**Do COVID-19 and crude oil prices drive the US economic policy uncertainty?**

Claudiu Tiberiu ALBULESCU[1,2*]

[1] Management Department, Politehnica University of Timisoara, 2, P-ta. Victoriei, 300006 Timisoara, Romania.

[2] CRIEF, University of Poitiers, 2, Rue Jean Carbonnier, Bât. A1 (BP 623), 86022, Poitiers, France.

**Abstract**

This paper investigates the effect of the novel coronavirus and crude oil prices on the United States (US) economic policy uncertainty (EPU). Using daily data for the period January 21 – March 13, 2020, our Autoregressive Distributed Lag (ARDL) model shows that the new infection cases reported at global level, and the death ratio, have no significant effect on the US EPU, whereas the oil price negative dynamics leads to increased uncertainty. However, analyzing the situation outside China, we discover that both new case announcements and the COVID-19 associated death ratio have a positive influence on the US EPU.

**1. Introduction**

[*] E-mail: claudiu.albulescu@upt.ro.

The outbreak of the new coronavirus (COVID-19) crisis monopolizes these days the worldwide public agendas. Originating in China (Hubei region), the COVID-19 affected over the last two months over 100,000 people and more than 100 countries. The World Health Organization (WHO), which daily monitors the COVID-19 figures since January 21, 2020, declared the coronavirus a pandemic. Although the spread of the virus started to decline, after the middle of February in China, the infection cases grew exponentially outside China. The European countries, but also the United States (US), are now severely touched. On the on hand, the COVID-19 triggers fear and anxiety in the society, nourished both by the daily reported new infection cases and by the increasing fatality ratio. On the other hand, the virus starts to affect the real economy, generating a crash on financial and commodity markets. Likewise, COVID-19 induces additional uncertainty in the economy, amplified by a delayed reaction of authorities, and by a lack of a clear strategy to fight against the disease.

This is also the case of the US authorities, which initially refused to take actions. President Trump declared that the situation is "very much under control" in the US, and that the scientists will find "soon" a solution to this problem. Nevertheless, the COVID-19 numbers triggered a strong reaction of US financial markets, first on February 28, 2020, and second, on March 13, 2020, immediately after Mr. Trump declaration according to which the coronavirus represents a "national emergency" issue. In addition, in the context of the coronavirus crisis, the Saudi Arabia decided to flood the market with oil and the international prices dropped over 20% on March 9, 2020. Against this background, we may ask if the coronavirus crisis influences the policy-induced economic uncertainty in the US. We tempt to respond to this question by analyzing the impact of new infection case announcements, and COVID-19 death-associated ratio, on the US economic policy uncertainty (EPU). We use daily data over the period January 21 – March 13, 2020, resorting to the WHO situation reports. We also test the impact of crude oil prices on the US EPU. Given that the US is an oil-dependent country, as one of the world largest producers, it is expected that a crash of crude oil price will amplify the economic uncertainty.

The relationship between EPU and oil prices has already been investigated. Chen et al. (2020) show that the impact of oil price shocks on EPU is positive at all frequencies, a result in line with Kang et al. (2017), but contrasting that reported by Antonakakis et al. (2014). At the same time, Ma et al. (2018) notice that EPU is important to forecast oil futures prices, whereas Aloui et al. (2016) show that EPU influences the oil price returns only in certain periods. Nevertheless, none of these studies focuses on the recent situation generated by the COVID-19 crisis. Therefore, we fill in this gap and test the impact of coronavirus numbers and West Texas Intermediate (WTI) prices on the US EPU (we use BRENT crude for robustness purpose). As far as we know, this is the first paper addressing the impact of the COVID-19 crisis on the US policy-induced economic uncertainty.

Only few works try to identify the causes underlying EPU (e.g. Chen et al., 2020; Duca and Saving, 2018). Indeed, most of existing studies address the impact of EPU on economic activity (Nyamela et al., 2019; Sahinoz and Cosar, 2018), financial volatility (Mei et al., 2018; Tiwari et al., 2019), bank valuation (He and Niu, 2018) or firm performance (Iqbal et al., 2019; Wu et al., 2020). Consequently, we add to the set of analyses investigating the drivers of EPU, with a focus on the officially reported coronavirus numbers.

## 2. Data

The WHO data show that China still is, for the moment, the most affected country in the world. However, the virus rapidly spreads in other countries like Italy, Spain, Iran or South Korea. Therefore, we posit that the impact of COVID-19 on the US EPU will be more important if the situation outside China degenerates. Figure 1 shows that the US EPU increases exponentially and seems to be positively correlated with the COVID-19 numbers.

Fig. 1. COVID-19 new infection cases and the US EPU

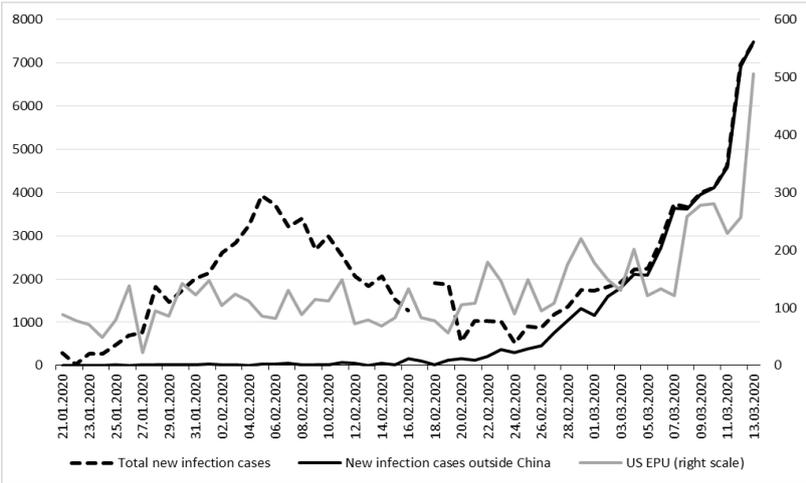

Figure 2 highlights that the death ratio has continuously increased, with a similar trend as that of the EPU index.

Fig. 2. COVID-19 death ratio and the US EPU

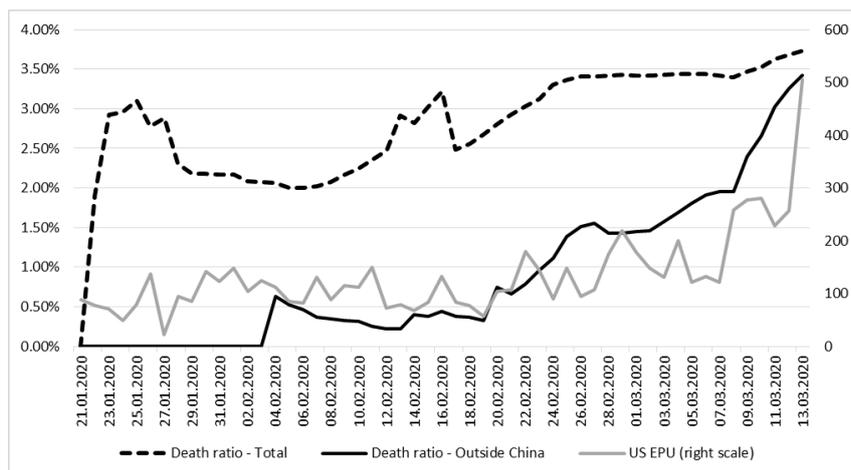

Our daily data are extracted from the WHO situation reports and refer to the new reported cases of infection and death ratio.[1] The EPU statistics come from Baker et al. (2016)[2], whereas the WTI crude oil prices are obtained from the US Energy Information Administration. All series are expressed in natural log.

## 3. Empirical investigation

### 3.1. Methodology

Given that our series are either I(0) and I(1) (Table 1), we use the ARDL model proposed by Pesaran et al. (2001) to investigate the relationship between the retained variables.

Table 1. ADF unit root test

|  | EPU | COVID-19$_{TNC}$ | COVID-19$_{NCOC}$ | COVID-19$_{TDR}$ | COVID-19$_{DROC}$ | WTI |
|---|---|---|---|---|---|---|
| Level | -3.224** | -3.692*** | 0.541 | -5.843*** | 0.079 | 1.276 |
| First difference | -11.69*** | -13.31*** | -7.413*** | -8.359*** | -6.946*** | -9.071*** |

Notes: (i) ***, ** and * means significance at 1%, 5% and 10%; (ii) the optimal lag selection is based on AIC information criterion; (iii) COVID-19$_{TNC}$ refers to total new infection cases, COVID-19$_{NCOC}$ shows the new cases reported outside China, COVID-19$_{TDR}$ is the total death ratio, COVID-19$_{DROC}$ is the death ratio outside China.

We therefore test the following specification that integrates the short-run adjustments into the long-run equilibrium:

---

[1] The WHO reports released at date "t" present data reported at "t-1", which are not final data. However, given the exponential increase of new infection cases, there is a delay of two days in the reporting activity. Therefore, to account for this delay, and to consider EPU's rapid reaction to coronavirus news, in our estimation we investigate the impact of numbers reflected in WHO reports released at date "t+1" on the EPU at date "t".
[2] Data are extracted on March 14, 2020 from http://www.policyuncertainty.com/us_daily.html.

$$\Delta EPU_t = c + \delta_{EPU} EPU_{t-1} + \delta_{COVID\text{-}19} COVID\text{-}19_t + \delta_{oil} Oil_{t-1} + \sum_{i=1}^{p} \alpha_i \Delta EPU_{t-i} +$$

$$\sum_{i=-1}^{p} \beta_i \Delta COVID\text{-}19_{t-i} + \sum_{i=0}^{p} \gamma_i \Delta Oil_{t-i} + \theta ECT_{t-i} + \varepsilon_t \qquad (1)$$

where: (i) $\Delta$ and $\delta$ are short- and long-run terms respectively (ii) "i" represents the maximum number of lags, (iii) the error correction adjustment term is denoted by ECT and the speed of adjustment is $\theta$, (iv) $\varepsilon$ is the error term.

### *3.2. Results*

We first validate the existence of the long-run relationship applying the bound test. For all tested models, the bound analysis indicates the existence of cointegration (Table 2).

Table 2. Bound test results

| Model specification | F-statistic | Critical values | | Conclusion |
|---|---|---|---|---|
| | | Lower bound (I(0)) | Upper bound (I(1)) | |
| COVID-19$_{TNC}$ | 10.54 | 3.10 | 3.87 | cointegration |
| COVID-19$_{NCOC}$ | 12.64 | 3.10 | 3.87 | cointegration |
| COVID-19$_{TDR}$ | 10.36 | 3.10 | 3.87 | cointegration |
| COVID-19$_{DROC}$ | 11.53 | 3.10 | 3.87 | cointegration |

Notes: (i) Critical values at 5% significance level; (ii) COVID-19$_{TNC}$ refers to total new infection cases, COVID-19$_{NCOC}$ shows the new cases reported outside China, COVID-19$_{TDR}$ is the total death ratio, COVID-19$_{DROC}$ is the death ratio outside China.

In the second step (Table 3), we present the results of the ARDL estimations for the four models we test, considering the total new infection cases (Model 1), the new cases reported outside China (Model 2), the total death ratio (Model 3) and the death ratio outside China (Model 4). The first model shows that at equilibrium (long run), the total new infection cases have no effect on the US EPU, but a decrease of crude oil price leads to a higher uncertainty, as expected. The short-run results show no impact of COVID-19 figures and WTI prices on the US EPU. However, the situation is different for Model 2, where the new infection cases reported outside China increase the uncertainty. At the same time, Model 4 shows that an increase of 1% in the logarithmic level of the death ratio recorded outside China, leads to an increase of 0.18% of the uncertainty index in the long run. This result remains robust when we use the BRENT crude as a *proxy* for the international oil prices.[3]

---

[3] The results can be provided upon request.

Table 3. Results of the ARDL specification

|  | Model 1: COVID-19$_{TNC}$ | Model 2: COVID-19$_{NCOC}$ | Model 3: COVID-19$_{TDR}$ | Model 4: COVID-19$_{DROC}$ |
|---|---|---|---|---|
| Long-run equation | | | | |
| COVID-19$_{t+1}$ | 0.040 | 0.068** | 0.057 | 0.182* |
| WTI$_t$ | -2.208*** | -1.426** | -2.401*** | -1.240* |
| c | 13.06*** | 10.00*** | 13.97*** | 9.431*** |
| Short-run equation | | | | |
| ΔEPU$_{t-1}$ | 0.456* | 0.481** | | |
| ΔCOVID-19$_{t+1}$ | | -0.010 | | |
| ΔCOVID-19$_{t-1}$ | | | 0.018 | |
| WTI$_{t-1}$ | | | -0.921 | -0.878 |
| ECT$_t$ | -0.947*** | -1.017*** | -0.958*** | -0.992*** |
| Tests on residuals | | | | |
| Serial correlation | NO | NO | NO | NO |
| ARCH effects | NO | NO | NO | NO |

Notes: (i) ***, ** and * means significance at 1%, 5% and 10%; (ii) Breusch-Godfrey LM test for serial correlation is used; (iii) ARCH effects for conditional heteroscedasticity (with 4 lags).

## 4. Conclusion

The rapid propagation of COVID-19 pandemic generates shock waves on the financial and commodity markets, as well as in the real economy. The depth of the new economic downturn will depend on the policy response to the coronavirus crisis. This paper investigates how COVID-19 official numbers (new infection cases and death ratio) affect the US EPU.

The findings show that the global COVID-19 numbers have no significant impact on the US EPU. Nevertheless, these results are largely influence by the situation reported in China, which seems to win the fight against the virus. When we assess the situation outside China, we clearly notice a positive influence of COVID-19 numbers on the US EPU. We therefore underline an amplification of COVID-19 risk to financial and real economy, generated by an increased, policy-induced economic uncertainty in the US.